\documentclass[12pt]{article}
\usepackage{graphicx,epsf}
\newcommand{\bea}{\begin{eqnarray}}
\newcommand{\eea}{\end{eqnarray}}

\newcommand{\gv}{\mbox{GeV}}
\newcommand{\mv}{\mbox{MeV}}

\newcommand{\be}{\begin{equation}}
\newcommand{\ee}{\end{equation}}
\newcommand{\ep}{\varepsilon}

\newcommand{\MSb}{$\overline{\rm MS}$ }

\newcommand{\MSbm}{\overline{\rm MS}}

\begin{document}

%%%%%%%%%%%%%%%%%%%%%%%%%%%%%%%%%%%%%%%%%%%%%%%%%%%%%%%%%%%%%%%%%%%%%%%%%%%%

\title{
\vskip-3cm{\baselineskip14pt
\centerline{\normalsize DESY~12-244\hfill}
\centerline{\normalsize HU-EP-12/54\hfill}
\centerline{\normalsize December 2012\hfill}}
\vskip1.5cm
On the difference between the pole and the \MSb masses of the top quark at the electroweak scale 
}

\author{
{\sc Fred Jegerlehner},
\\
\\
{\normalsize Humboldt-Universit\"at zu Berlin, Institut f\"ur Physik,}\\ 
{\normalsize  Newtonstra\ss e 15, 12489 Berlin, Germany}\\
{\normalsize Deutsches Elektronen-Synchrotron (DESY),}\\ 
{\normalsize Platanenallee 6, 15738 Zeuthen, Germany}
\\
\\
{\sc Mikhail~Yu.~Kalmykov}\thanks{On leave of absence from
Joint Institute for Nuclear Research, 141980 Dubna (Moscow Region), Russia.},
{\sc Bernd~A.~Kniehl}
\\
\\
{\normalsize II. Institut f\"ur Theoretische Physik, Universit\"at Hamburg,}\\
{\normalsize Luruper Chaussee 149, 22761 Hamburg, Germany}
}

\date{}

\maketitle
%%%%%%%%%%%%%%%%%%%%%%%%%%%%%%%%%%%%%%%%%%%%%%%%%%%%%%%%%%%%%%%%%%%%%%%%%%%%%%%
\abstract{
We argue that for a Higgs boson mass $M_H \sim 125$~GeV, as suggested by
recent Higgs searches at the LHC, the inclusion of electroweak
radiative corrections in the relationship between the pole and \MSb
masses of the top quark reduces the difference to about 1~GeV.
This is relevant for the scheme dependence of electroweak observables,
such as the $\rho$ parameter, as well as for the extraction of the top quark
mass from experimental data.
In fact, the value currently extracted by reconstructing the invariant mass of
the top quark decay products is expected to be close to the pole mass, while
the analysis of the total cross section of top quark pair production yields a
clean determination of the \MSb mass. 

\medskip

}

\newpage

%=====================================================================
\renewcommand{\thefootnote}{\arabic{footnote}}
\setcounter{footnote}{0}

%%%%%%%%%%%%%%%%%%%%%%%%%%%%%%%%%%%%%%%%%%%%%%%%%%%%%%%%%%%%%%%%%%%%%%%%%
%
%%%%%%%%%%%%%%%%%%%%%%%%%%%%%%%%%%%%%%%%%%%%%%%%%%%%%%%%%%%%%%%%%%%%%%%%%%%%

%=====================================================================
\renewcommand{\thefootnote}{\arabic{footnote}}
\setcounter{footnote}{0}
%%%%%%%%%%%%%%%%%%%%%%%%%%%%%%%%%%%%%%%%%%%%%%%%%%%%%%%%%%%%%%%%%%%%%%%%%
%\section{Introduction}
%\setcounter{equation}{0}
%%%%%%%%%%%%%%%%%%%%%%%%%%%%%%%%%%%%%%%%%%%%%%%%%%%%%%%%%%%%%%%%%%%%%%%%%%%%

%=====================================================================

%%%%%%%%%%%%%%%%%%%%%%%%%%%%%%%%%%%%%%%%%%%%%%%%%%%%%%%%%%%%%%%%%%%%%%%%%%%%%
\section{Introduction}
For the precise understanding of the relationship between running and
pole masses of particles, within the framework of the Standard Model
(SM) of electroweak (EW) and strong interactions, it is mandatory to use
the full SM renormalization group (RG) equations.  In this paper, we
focus in particular on the top quark mass.  In published results,
commonly only the QCD corrections are applied, but also the
corresponding EW corrections are important. Here we
discuss the EW contributions to the SM RG equations and the related
matching conditions and their numerical significance for the pole
mass. The relevant corrections have been derived for the top quark in 
Refs.~\cite{Yukawa:1,pole-top,Jegerlehner:2003sp}. 
Assuming the particle recently discovered at the CERN LHC \cite{ATLAS} 
to be the SM Higgs boson, it is possible
to specify the corrections numerically. As we know the top quark, like
no other quark, is accessible to perturbative predictions by virtue of
its very large mass and small width, which let the top quark decay before it
can form hadrons.

Since free quarks are not observable in nature, their masses
primarily are Lagrangian parameters which parametrize the chiral
symmetry breaking in terms of masses as required by observation,
mainly by the observed mass spectrum of the hadronic states, which
consist of permanently confined quarks and gluons. In any case, quark
masses are needed as input parameters for calculations of SM
predictions~\cite{scheme} and must be tuned to account for
corresponding mass effects in hadronic reactions. The most frequently
used definitions of mass are the pole and \MSb ones, which
for quarks both are formal definitions. They both are popular because
of their simple access in perturbation theory. One should note that
the \MSb scheme is intrinsically only defined in the perturbative
approach.

Applying dimensional regularization~\cite{dimreg} and the $\ep=(4-d)/2$
expansion, the RG functions are uniquely defined, order by order in
perturbation theory, by the ultraviolet (UV) properties of the model, represented by
the $1/\ep$ counterterms~\cite{RG}. In order to determine the value of a
running mass at some scale, the matching condition between the running
mass and some observable has to be evaluated (see
e.g.\ Ref.~\cite{Passarino}). Since the SM includes both EW and
QCD type UV singularities, the corresponding RG equations have to take
into account both, too.

The pole mass is a well defined quantity within perturbation
theory. It is related to the pole of the renormalized propagator in
the complex energy plane. The position of the pole is a gauge
invariant and infrared finite quantity~\cite{pole:QCD,pole:SM}.  A
shortcoming is the fact that the pole mass suffers from renormalon
contributions, which worsen the convergence properties of the
perturbative expansion. The corresponding uncertainty is of the order
of $\Lambda_{\rm QCD} \sim 200~\mv$~\cite{renormalon,top-renormalon}, which is
not too large for a particle as heavy as the top quark, but leads to an
intrinsic limitation of the possible precision.
The top quark being a colored object,
the pole of its propagator is not an observable per se, although it
seems that the color singlet recombination via gluonic strong-interaction
effects does not affect the location of the top quark propagator
pole very much. These problems and deficiencies have triggered many
discussions about the accuracy of the top quark mass and its
extraction from experimental data, and actually other mass definitions
which look to be closer to observable quantities have been worked
out~\cite{Hoang,Alioli:2012es}. Usually, alternative masses are nevertheless
converted into pole and/or \MSb masses, which thus both remain useful
concepts, and their relationship remains of primary interest. However,
up to now, mostly QCD corrections have been included in the conversion
between pole and \MSb masses of the top quark. In this note, we shall
discuss how to account for the EW contributions and evaluate their 
size.
We shall denote a pole mass by capital $M$ and a \MSb mass by lowercase
$m$ in the following.
%%%%%%%%%%%%%%%%%%%%%%%%%%%%%%%%%%%%%%%%%%%%%%%%%%%%%%%%%%%%%%%%%%%%%%%%%%%%%%%%%%%%%
%%%%%%%%%%%%%%%%%%%%%%%%%%%%%%%%%%%%%%%%%%%%%%%%%%%%%%%%%%%%%%%%%%%%%%%%%%%%%%%%%%%%%
\section{Running masses in the SM}

The first systematic inclusion of the EW corrections in the
definition of the running mass of a fermion has
been achieved in Ref.~\cite{Yukawa:1}. By including all self-energy diagrams
(including tadpoles), one obtains a gauge invariant relation between
pole and bare masses~\cite{FJ81}. By applying minimal subtraction to
the UV counterterms of this relation, the one-loop relation between a
\MSb mass $m_f$ and the corresponding pole mass $M_f$, as well as the
threshold relation $\delta_{f,\alpha}$ between the corresponding
Yukawa coupling $y_f(\mu^2)$ and $M_f$, have been derived. 
In this approach, care has to be exercised, especially at the multiloop level,
to include all the contributing diagrams including tadpoles, while it
is not sufficient to select gauge invariant subsets.
As an illustrative example, we mention the $O(\alpha \alpha_s)$ mixing
contributions to the pole masses of quarks. The definition, via a ``gauge
invariant set of diagrams including tadpole contributions", was
complemented in Ref.~\cite{pole:1} by a theorem about the interrelation
between the RG functions for the massive parameters
(masses of particles, as well as the Fermi constant) calculated in the
broken phase of the SM with RG functions of parameters of the unbroken
phase of the SM, in accord with the expectation that spontaneous
symmetry breaking does not affect the UV structure of the SM. In other
words, the EW UV counterterms in the broken phase of the SM can be
obtained in terms of the UV counterterms in the unbroken phase.\footnote{A
different theorem states that tadpole terms, which are absent in the symmetric
phase, drop out from observable quantities. However, if one omits
tadpole terms in relations between bare and renormalized quantities,
as frequently done in SM calculations
\cite{Passarino,Faisst:2003px,Kuhn:2006vh,martin}, one not
only looses a manifestly gauge invariant relationship between the bare and the
renormalized theory, also the UV structure is not preserved and
one does not get the same RG equations. In fact, tadpoles are related
to quadratic divergences which show up in the renormalization of the
mass parameter $m^2$ of the Higgs potential in the symmetric
phase.} The above-mentioned theorem has been verified explicitly by a
two-loop analysis of the UV counterterms evaluated in the broken phase
of SM~\cite{pole-top,Jegerlehner:2003sp,pole:1}. This approach gives rise to the same set
of quark self-energy Feynman diagrams~\cite{Yukawa:2} as well as to an
equivalent definition~\cite{Yukawa:3} of the threshold
relations~\cite{Yukawa:1,Sirlin}.

Before we proceed, let us remind the reader of some basic definitions needed
for the following discussion. Applying dimensional regularization~\cite{dimreg} in the
broken phase, the SM UV counterterms for the quark masses in the \MSb
scheme have the following form:
\begin{equation}
m_{q,{\rm bare}} = m_q(\mu^2)
\Biggl[ 
1 
+ \alpha_s \sum_{i=0} \alpha_s^i \sum_{k=1}^{i+1} \frac{\delta Z_{\alpha_s}^{(i,k)}}{\ep^k}
+ \alpha \sum_{i,j=0} \alpha^i \alpha_s^j \sum_{k=1}^{i+j+1} \frac{\delta Z_\alpha^{(i,j,k)}}{\ep^k}
\Biggr]\;.
\label{bare}
\end{equation}
The first series in this relation corresponds to the QCD corrections,
the second one to the EW contribution mixed in higher orders
with QCD. In accordance with 't Hooft's prescription~\cite{RG}, the
quark mass anomalous dimension, defined by
\begin{eqnarray}
\mu^2 \frac{d}{d \mu^2} \ln m_q^2  
= \gamma_q(\alpha_s,\alpha) 
= 
\left[ 
\alpha_s \frac{\partial}{\partial \alpha_s }
+ 
\alpha \frac{\partial}{\partial \alpha }
\right]
\Biggl[ 
\alpha_s \sum_{i=0} \alpha_s^i \delta Z_{\alpha_s}^{(i,1)}
\!+\! 
\alpha  \sum_{i,j=0} \alpha^i \alpha_s^j \delta Z_\alpha^{(i,j,1)}
\Biggr] 
\;,
\label{running-mass}
\end{eqnarray}
can be split into two parts: the QCD and EW contributions  
%\begin{eqnarray}
$
\gamma_q(\alpha_s,\alpha)
= 
\gamma_q^{\rm QCD} 
+ 
\gamma^{\rm EW}_q \;,
%\end{eqnarray}
$ where $\gamma_q^{\rm QCD}$ includes all terms which are proportional to
powers of $\alpha_s$ only and $\gamma^{\rm EW}_q$ includes all other terms
proportional to at least one power of $\alpha$, and beyond one loop
multiplied by further powers of $\alpha$ and/or $\alpha_s$. We call
$\gamma_t^{\rm QCD}$ the QCD anomalous dimension and $\gamma^{\rm EW}_t$ the
EW one. As has been shown in Ref.~\cite{pole:1}, the
EW contribution to the fermion anomalous dimension
$\gamma^{\rm EW}_t$ in the $\overline{\rm MS}$ scheme may be written in
terms of RG functions of parameters in the unbroken phase of the SM as
\begin{eqnarray}
\gamma^{\rm EW}_t & = & \gamma_{y_t} + \frac{1}{2} \gamma_{m^2} - \frac{1}{2} \frac{\beta_\lambda}{\lambda} \;,
\label{SM<->F}
\end{eqnarray}
where 
$
\gamma_{m^2} = \mu^2 \frac{d}{d \mu^2} \ln m^2
$,
%%%%%%%%%%%%%%%%%%%%%%%%%%%%%%%%%%%%%%%%%%%%%%%%%%%%%%%%%%%%%
$
\beta_\lambda = \mu^2 \frac{d}{d \mu^2} \lambda
$,
%%%%%%%%%%%%%%%%%%%%%%%%%%%%%%%%%%%%%%%%%%%%%%%%%%%%%%%%%%%%%%
$
\gamma_{y_q} = \mu^2 \frac{d}{d \mu^2} \ln y_q 
$,
$y_q$ is the Yukawa coupling of quark $q$, 
and  $m^2$ and $\lambda$ are the parameters of the scalar potential
$
V = \frac{m^2}{2} \phi^2 + \frac{\lambda}{24} \phi^4$.

It has also been shown \cite{JK2004} that the coefficients of
the higher poles in $\ep$ in the mass counterterms (\ref{bare}) in
the broken phase are uniquely determined by the lower-order coefficients
and the RG functions defined by Eq.~(\ref{SM<->F}).

The RG equation for the square of the Higgs vacuum expectation value (VEV)
$v(\mu^2)$ follows from the RG equations for masses and
massless coupling constants and reads
\begin{eqnarray}
\mu^2 \frac{d}{d \mu^2} v^2(\mu^2)
& \!\!=\!\!\! &4\, \mu^2 \frac{d}{d \mu^2} \left[\frac{m_W^2(\mu^2)}{g^2(\mu^2)} \right]
=4\, \mu^2 \frac{d}{d \mu^2} \left[\frac{m_Z^2(\mu^2)-m_W^2(\mu^2)}{g'^2(\mu^2)} \right]
\nonumber \\ 
& \!\!=\!\!\! &3\, \mu^2 \frac{d}{d \mu^2} \left[\frac{m_H^2(\mu^2)}{\lambda(\mu^2)} \right]
=2\, \mu^2 \frac{d}{d \mu^2} \left[\frac{m_f^2(\mu^2)}{y^2_f(\mu^2)} \right] 
=
v^2(\mu^2) \left[\gamma_{m^2}  - \frac{\beta_\lambda}{\lambda} \right],
\label{vev}
\end{eqnarray}
where $g^\prime$ and $g$ are the $U(1)_Y$ and $SU(2)_L$ gauge couplings,
respectively, and we assume the running of $g$ and $g^\prime$ as well as
of $y_t$ and $\lambda$ to be the same in the broken and the unbroken phases
\cite{Yukawa:3,degrassi,Moch,Mihaila:2012fm,Bednyakov:2012rb,Chetyrkin:2012rz,Masina:2012tz,Hamada:2012bp}.
Since the relation\footnote{By $G_F$ we denote a generic Fermi
constant, by $G_\mu$ the physical on-shell one, and by
$G_F^{\MSbm}$ the \MSb variant.} $G_F=\frac{1}{\sqrt{2}\,v^2}$ is valid
for bare as well as for on-shell parameters, the RG equation for the
\MSb version of the running Fermi constant follows from $G^{\MSbm}_F(\mu^2)=\frac{1}{\sqrt{2}\,v^2(\mu^2)}$.
The corresponding anomalous dimension $\gamma_{G_F}$ of $G^{\MSbm}_F$
in then given by
\begin{eqnarray}
\gamma_{G_F}= \mu^2 \frac{d}{d \mu^2} \ln\,G^{\MSbm}_F(\mu^2)=-
\mu^2 \frac{d}{d \mu^2} \ln\,v^2(\mu^2) =- \left[\gamma_{m^2}  - \frac{\beta_\lambda}{\lambda} \right],
\label{GFRG}
\end{eqnarray}
i.e., by minus the anomalous dimension of $v^2$.

We note that the anomalous dimension of $v^2(\mu^2)$ defined by
Eq.~(\ref{vev}) via diagrammatic calculations differs from the anomalous
dimension of the scalar field as obtained in the effective-potential
approach \cite{RG-SM}.

The RG equations (\ref{running-mass}) have to be complemented by
matching conditions between pole and running masses, which we may write in the form
\begin{eqnarray}
M_t - m_t(\mu^2) = 
m_t(\mu^2) \sum_{j=1} \left( \frac{\alpha_s(\mu^2)}{\pi} \right)^j \rho_j 
+ 
m_t(\mu^2) \sum_{i=1;j=0} \left( \frac{\alpha(\mu^2)}{\pi} \right)^i  \left( \frac{\alpha_s(\mu^2)}{\pi} \right)^j r_{ij} \;. 
\label{matching}
\end{eqnarray}
The QCD corrections 
$
\rho_j 
$ were calculated in
Refs.~\cite{ms-pole:qcd,ms-pole:qcd2,ms-pole:qcd3} up to $j=3$, while
the $O(\alpha)$ and $O(\alpha \alpha_s)$ corrections $r_{10}$ and
$r_{11}$, respectively, are available in analytic form from
Refs.~\cite{Yukawa:1,pole-top,Jegerlehner:2003sp}.  The
$O(\alpha\alpha_s)$ result for $r_{11}$ with tadpoles dropped was also
evaluated using asymptotic expansions in Ref.~\cite{Eiras:2005yt} and
numerical agreement with Refs.~\cite{pole-top,Jegerlehner:2003sp} was
found after subtraction of the tadpoles. The leading part of the
$O(G_\mu M_t^2\alpha_s)$ contribution to $r_{11}$ was confirmed in
Ref.~\cite{Faisst:2004gn} after including the tadpole contribution in
the result of Ref.~\cite{Faisst:2003px}.  The correction $r_{12}$ has
been evaluated in Ref.~\cite{martin} in the gaugeless-limit
approximation of the SM.

%%%%%%%%%%%%%%%%%%%%%%%%%%%%%%%%%%%%%%%%%%%%%%%%%%%%%%%%%%%%%%%%%%%%%%%%%%%%%%%%%%%%%%%%%%%%%%%%%
\section{Behavior of the RG equations at low and high energies}
Let us analyze the behavior of the full SM RG equation for a quark mass in the \MSb scheme
\begin{equation}
\mu^2\, \frac{d}{d\mu^2} \: \ln m_f(\mu^2) 
%= 
%\mu^2\, \frac{d}{d\mu^2} \: \ln \left( y_f(\mu^2) v(\mu^2) \right) 
%= 
%\mu^2\, \frac{d}{d\mu^2} \: \ln \frac{y_f(\mu^2)}{\sqrt{G_F(\mu^2)}}
= \gamma_q^{\rm QCD} + \gamma_{y_f} - \frac{1}{2} \gamma_{G_F} \;,
\label{RG}
\end{equation}
in which the EW part follows from Eqs.~(\ref{SM<->F}) and (\ref{vev}). 
Let us consider the low-energy limit first. In the weak sector of the
SM, there is no decoupling because masses and couplings are
interrelated by the Higgs mechanism. So ``decoupling by hand'' as
usually applied in QCD by considering an effective `$n_f$ active
flavors' QCD to be matched at successive flavor thresholds, and which
can be applied to QED as well, does not make sense in the weak sector.
Note that there is no decoupling for the $W$ and $Z$ bosons: the limit $M_W\to
\infty$ can be achieved by letting $g \to \infty$ or $v
\to \infty$ or both. In nature, only the limit $g \to \infty$ leads to
the observed low-energy limit of the effective four-fermion theory
with $\sqrt{2}\,G_\mu=1/v^2$ fixed, by nuclear $\beta$ decay
etc. This obviously is a non-decoupling effect. In contrast to QED or
QCD, the low-energy effective theory (obtained after elimination of the
heavy state) is a non-renormalizable one exhibiting a completely wrong
high-energy behavior. So, in general, ``decoupling by hand,'' as it is
commonly utilized in \MSb-parametrized QCD, is not very sensible
when the Appelquist-Carazzone theorem~\cite{Appelquist:1974tg} does not apply.

Nevertheless, in calculations of EW radiative corrections for
LEP processes, covering scales up to 200~GeV, the standard on-shell
parametrization in terms of the most precise parameters $\alpha$,
$G_\mu$, and $M_Z$ (besides the fermion and Higgs-boson masses) reveals that, 
while $\alpha$ is running strongly, keeping $G_\mu$ as scale
independent\footnote{This assertion has been checked
experimentally by comparing the standard low-energy quantity
$G_\mu$ determined via the  muon lifetime $\tau_\mu=1/\Gamma_\mu(\mu \to
e\bar\nu_e\nu_\mu)$ versus the corresponding effective coupling extracted
from the leptonic $W$-boson decay rate
$\hat{G}_\mu=12\pi \Gamma_{W\ell\nu}/(\sqrt{2}M_W^3)$, which
involves $W$-boson mass scale observables only. The fact that
$\hat{G}_\mu\approx G_\mu$ with good accuracy is not surprising because the tadpole corrections, 
which potentially lead to substantial
corrections, are absent in relations between
observable quantities as we know.} 
provides an excellent parametrization in terms of $\alpha(M_Z^2)$, $G_\mu$
and $M_Z$ for LEP observables. The latter parametrization incorporates
the leading-logarithmic resummation as effectuated by the RG. Usually the
scale insensitivity of an effective $G_{F}$ is ``explained'' by a
``decoupling by hand'' argument via inspection of the one-loop RG equation
\begin{equation}
\mu^2\,\frac{d G^{\MSbm}_F}{d \mu^2}=\frac{G^{\MSbm}_F}{8\pi^2\sqrt{2}}\biggl\{
\sum_{f}
\left(m_f^2-4\,\frac{m_f^4}{m_H^2}\right) -3\,M_W^2+6\,\frac{M_W^4}{m_H^2} -
\frac{3}{2}M_Z^2+3\,\frac{M_Z^4}{m_H^2}
+\frac{3}{2}m_H^2 \biggr\}\,,
\label{GmuRG}
\end{equation}
which follows from the counterterm given first in Ref.~\cite{FJ81}. If we only
sum terms with $m_f<\mu$, there is effectively no running (because of the
smallness of the light-fermion masses) before $M_W$, $M_Z$, $M_H$, and $M_t$
come into play.

As mentioned earlier, ambiguities enter if we are to represent
predictions in terms of the not-so-physical \MSb parameters.\footnote{The
\MSb parameters other than $v(\mu^2)$ (i.e. \MSb gauge, Yukawa, and Higgs
self couplings) 
are likely the most natural parameters in the unbroken phase of the SM,
where an $S$ matrix does not exist due to infrared problems. Other
renormalization schemes that can be applied in this case include
the MOM-type schemes, which are, however, gauge dependent.} On
phenomenological grounds, as $G_F$ has been measured to agree at the
$M_Z$ scale with its low-energy version $G_\mu$ and because Yukawa
couplings run as they do in the symmetric phase, below of the EW
scale, one may define effective light-fermion masses to run via
their Yukawa couplings only:
\begin{equation}
\hat{m}_f(\mu^2) =  2^{-3/4}G_\mu^{-1/2}y_f(\mu^2) \;.
\label{effhat}
\end{equation}
As the Yukawa couplings
$y_f(\mu^2)$ are not affected by the Higgs mechanism, the EW
corrections to the Yukawa couplings are free of
tadpoles~\cite{Yukawa:1,Yukawa:2} and/or quadratic divergences.  Since
real physical observables are also free of tadpole contributions, this
property is an additional argument why Eq.~(\ref{effhat}) is a good
candidate for the evaluation of the EW contributions to the ratio
between pole and \MSb masses of lighter quarks, such as the bottom and
charm quarks (see also the discussion in Ref.~\cite{willey}). In
short, fermion masses and Yukawa couplings have equivalent RG
evolutions as long as $G_F$ or, equivalently, $v$ can be taken not to
be running, so that one may identify $G_F^{\mbox{\MSb}}(\mu^2)=G_\mu$.
Alternatively, and more
consequently concerning the decoupling issue, 
the proper \MSb definition of a running fermion mass is
\begin{equation}
m_f(\mu^2) =  2^{-3/4}\,\left(G^{\MSbm}_F(\mu^2)\right)^{-1/2}y_f(\mu^2) \;,
\label{eff}
\end{equation}
where $G^{\MSbm}_F(\mu^2)$ and $m_f(\mu^2)$ are solutions of
Eq.~(\ref{GFRG}) and (\ref{RG}), respectively. For the \MSb top quark mass, 
we consequently advocate to utilize Eq.~(\ref{eff}), which
among others includes the tadpole contributions. Note that
the difference between Eqs.~(\ref{effhat}) and (\ref{eff}) is
particularly significant for the top quark. As both versions are
gauge invariant by definition, the difference is not just dropping
the tadpole terms or not.

The running of $G^{\MSbm}_F$ definitely starts at about $\mu \sim
2M_W$,\footnote{As the on-shell version of $G_F$ at the $Z$-boson mass scale
can be identified with $G_\mu$, it is justified to match 
$G^{\MSbm}_F$ with $G_\mu$ at the scale $M_Z$.} when
the scale of a process exceeds the masses of the weak gauge
bosons. Since the top quark is the heaviest particle in the SM, at
least here the ``decoupling by hand'' prescription becomes obsolete,
and we have to take full SM parameter relations as they are.

One of the most well-known non-decoupling effects related to the top quark is
the EW $\rho$ parameter $\rho_{\rm eff}(0)=G_{\rm NC}/G_{\rm CC}(0)$,
where $G_{\rm CC}(0)$ is the Fermi coupling $G_F=G_\mu$ and $G_{\rm
NC}$ the low-energy effective axial-vector $Z$-boson coupling to
fermions. As is well known, in the SM we have
\begin{equation}
\rho=1+\frac{N_cG_\mu}{8\pi^2\sqrt{2}}
\left(m_t^2+m_b^2-\frac{2m_t^2m_b^2}{m_t^2-m_b^2}\ln\frac{m_t^2}{m_b^2}\right)
\approx 1+\frac{N_cy_t^2}{32\pi^2},
\end{equation}
which measures the weak-isospin breaking
by the Yukawa couplings of the heavy fermions at zero momentum. Within
the SM, this quantity is strongly constrained by LEP data, and, in spite
of the fact that the top quark was by far too heavy to be produced at LEP,
the top quark contribution and indirectly the top quark mass have been
constrained
by LEP data. Actually, a first strong indication of a heavy top quark had been
found earlier by the ARGUS experiment, which discovered,
unexpectedly, a substantial $B\overline{B}$ oscillation (in the SM enhanced
by a contribution $\propto y_t^2$), which turned out to be much
larger than anticipated before. So recipes like ``decoupling by hand''
make no sense to be applied to the weak sector of the SM, as heavy-particle
effects definitely cannot be renormalized away.

For large values of $\mu^2$, the behavior of the running Fermi
constant $G_F(\mu^2)$
is defined by the Higgs self-coupling and the sign of its
beta-function $\beta_\lambda$:
\begin{equation}
\mu^2 \frac{d}{d \mu^2} \ln G_F(\mu^2)  \sim  \frac{\beta_\lambda(\mu^2)}{\lambda(\mu^2)} \;.
\end{equation}
The detailed perturbative analysis of the r.h.s. of this equation was performed 
recently
%\cite{Yukawa:3,degrassi}
(see Refs.~\cite{Yukawa:3,degrassi,Moch,Mihaila:2012fm,Chetyrkin:2012rz,Masina:2012tz}) 
and reveals that the beta function $\beta_\lambda$ is negative up to a
scale of about $10^{17}~\gv$, where it changes sign. Above the zero of
$\beta_\lambda$, the effective coupling starts to increase again, and
the key question is whether at the zero of the beta function the
effective coupling is still positive. In the latter case, it will remain
positive
although small up to the Planck scale. In any case, at moderately high
scales where $\beta_\lambda <0$, and provided that $\lambda$ is still
positive, the following behavior is valid for the Fermi constant:
\begin{equation}
\left. G_F(\mu^2) \right|_{\mu^2 \to \infty} 
\sim  \left(\mu^2\right)^\frac{\beta_\lambda(\mu^2)}{\lambda(\mu^2)} 
\to 0 \;, 
\label{RG-PLANK}
\end{equation}
being decreasing, which means that $v^2(\mu^2)$ is increasing at these
scales (where $\beta_\lambda <0$ and $\lambda>0$). The
analysis of Ref.~\cite{degrassi} finds that $\lambda$ turns negative (unstable
or meta-stable Higgs potential) before the beta function reaches its
zero. This may happen at rather low scales, at around $10^{10}~\gv$. In
this case, we would get an infinite Higgs vacuum expectation value far below
the Planck scale as an essential singularity.
Given the present uncertainty in the value of $M_t$, there is a good chance
that $\lambda$ remains positive up to the zero of the beta function and as
a consequence up to the Planck scale~\cite{Yukawa:3, Moch}. Then
$G_F(\mu^2)$ would start to increase again, and $v(\mu^2)$ would
start to decrease but remain finite (about $685~\gv$) at the Plank
scale, implying that all effective masses stay bounded. The
effectively massless symmetric phase of the SM would then be obtained
at high energies by the fact that mass effects are suppressed for
dimensional reasons: according to the RG, for a vertex function under a
dilatation of all momenta, $\{p_i\}
\to \{\kappa p_i\}$, up to the overall dynamical dimension and
wave-function renormalizations, the result is given by replacing $g_i \to
g_i(\kappa)$ and $m_i \to m_i(\kappa)/\kappa$ at fixed $\{p_i\}$ and
renormalization scale $\mu$. I.e.\ provided that $m(\kappa)/\kappa \to 0$ as
$\kappa \to \infty$, the high-energy asymptotic effective theory is
effectively massless as expected in the symmetric phase.

%%%%%%%%%%%%%%%%%%%%%%%%%%%%%%%%%%%%%%%%%%%%%%%%%%%%%%%%%%%%%%%%%%%%%%%%%%%%%%%%%%%%%%%%%%
%%%%%%%%%%%%%%%%%%%%%%%%%%%%%%%%%%%%%%%%%%%%%%%%%%%%%%%%%%%%%%%%%%%%%%%%%%%%%%%%%%%%%%%%%%
%%%%%%%%%%%%%%%%%%%%%%%%%%%%%%%%%%%%%%%%%%%%%%%%%%%%%%%%%%%%%%%%%%%%%%%%%%%%%%%%%%%%%%%%%%

\boldmath
\section{Numerical result for $m_t-M_t$}
\unboldmath

In the previous section, we have presented the arguments, why
decoupling does not apply in the EW sector, in particular not to the 
top quark mass effects. In this section, we will check how significant
the EW contribution to matching and running of the top quark mass is.  For
that purpose, the inverse of the relation (\ref{matching}), $m_t(\mu^2)$
as a function of the pole mass $M_t$, is required (see Eq.~(5.54) in
Ref.~\cite{pole-top}).  For the numerical analysis, we adopt the following
values for the input parameters~\cite{pdg}:
\begin{eqnarray}
M_Z &=& 91.1876(21)~\gv, 
\quad 
M_W = 80.385(15)~\gv,
\quad 
M_t = 173.5(1.0)~\gv,\footnotemark
\nonumber \\ 
G_F &=&1.16637\times 10^{-5}~\gv^{-2}, 
\quad
\alpha^{-1} = 137.035999,\quad 
\alpha_s^{(5)}(M_Z^2) = 0.1184(7).\quad
\label{params}
\end{eqnarray}
\footnotetext{The values of the top quark mass quoted
by the experimental collaborations correspond to parameters in Monte
Carlo event generators in which, apart from parton showering, the partonic
subprocesses are calculated at the tree level, so that a rigorous theoretical
definition of the top quark mass is lacking \cite{pdg,Abazov:2011pta}.
For definiteness, we take the value from
Ref.~\cite{pdg} to be the pole mass $M_t$.}
Furthermore, we take the effective fine-structure constant at the $Z$
boson mass
scale to be $\alpha^{-1}(M_Z^2) = 127.944$. All light-fermion
masses $M_f~(f\neq t)$ give negligible effects and do not play any
role in our consideration.
%For our purpose, it suffices to include the
%one-loop runnings of the strong-coupling and electromagnetic
%fine-structure constants. For scales above the $W$ boson mass, but
%below the top quark mass, we then have
%\begin{eqnarray}
%\alpha_s(\mu^2) = \frac{\alpha_s(M_Z)}{1 \!+\!  \frac{23}{12}\,  \frac{\alpha_s(M_Z)}{\pi}\, \ln \frac{\mu^2}{M_Z^2} } \;, 
%\quad
%\alpha(\mu^2) = \frac{\alpha(M_Z)}{1 \!-\! \frac{11}{12}\,
%\frac{\alpha(M_Z)}{\pi}\, \ln \frac{\mu^2}{M_Z^2}} \;.
%\label{1-loop}
%\end{eqnarray}
Up to the three-loop order, the QCD relation between the running and pole
masses is given by (see Eq.~(12) in Ref.~\cite{ms-pole:qcd3})
\begin{eqnarray}
\left\{ m_t(M_t^2) - M_t \right\}_{\rm QCD}&=& 
M_t 
\Biggl[ 
- \frac{4}{3}\, \frac{\alpha_s^{(6)}(M_t^2)}{\pi}\, 
- 9.125\left( \frac{\alpha_s^{(6)}(M_t^2)}{\pi} \right)^2
\nonumber\\
&&{}- 80.405 \left( \frac{\alpha_s^{(6)}(M_t^2)}{\pi} \right)^3    
\Biggr] \;.
\end{eqnarray}
%Taking into account the running of $\alpha_s(\mu^2)$ from the scale $M_Z$ to
%$M_t$
%, where it suffices to use the one-loop approximation given above,
Using $\alpha_s^{(6)}(M_t^2)=0.1079(6)$ \cite{Kniehl:2012rz}, which follows from
the value of $\alpha_s^{(5)}(M_Z^2)$ in Eq.~(\ref{params}) via four-loop
evolution and three-loop matching \cite{Chetyrkin:1997sg},
we obtain the numerical result  
\begin{eqnarray}
\left\{ m_t(M_t^2) - M_t \right\}_{\rm QCD}= 
 - 7.95  ~\gv 
 - 1.87  ~\gv
 - 0.57  ~\gv
 = -10.38 ~\gv \;.
\label{qcd:num}
\end{eqnarray}
%The error due to this approximation is so small that it affects our numbers
%only below the accuracy to which they are presented.
%Including the available higher-order corrections to the running of
%$\alpha_s(\mu^2)$ would shift the result in Eq.~(\ref{qcd:num}) by less than
%{\bf\Huge ???}.
%%%%%%%%%%%%%%%%%%%%%%%%%%%%%%%%%%%%%%%%%%%%%%%%%%%%%%%%%%%%%%%%%%%%%%%%%%%%%%%%%%
\begin{figure}[t]
\begin{center}
\centerline{\vbox{\epsfysize=100mm \epsfbox{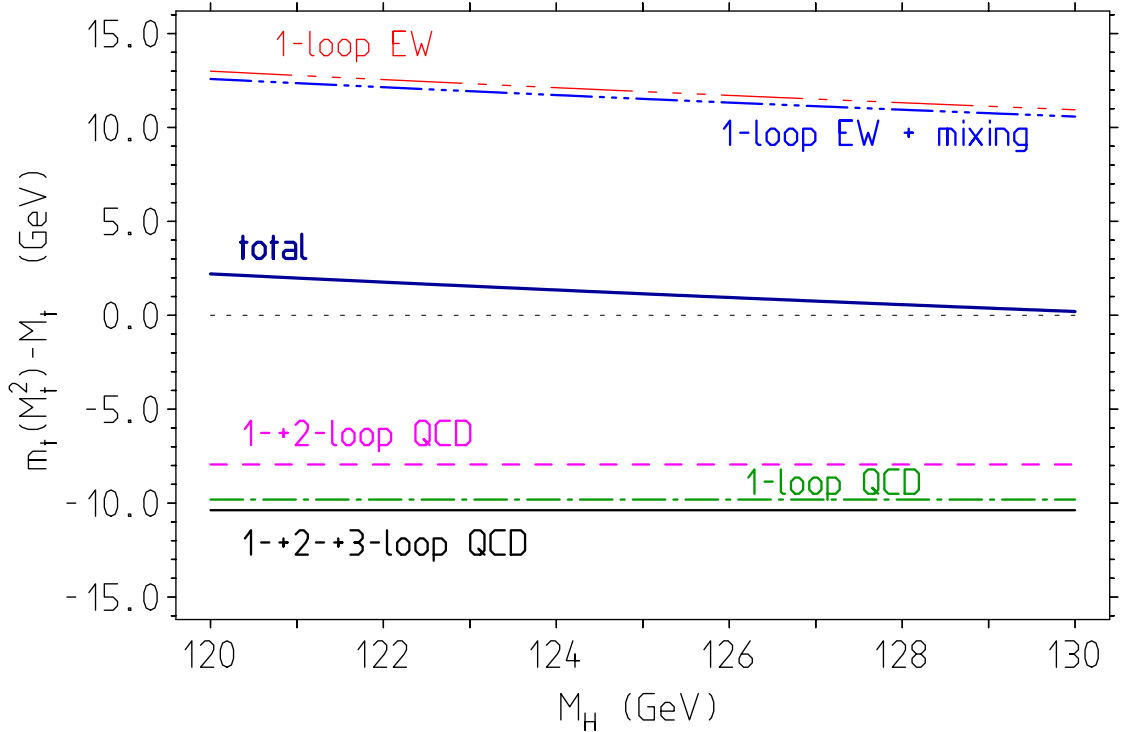}}}
%\centerline{\vbox{\epsfysize=100mm \epsfbox{pole-ms.eps}}}
\caption{
\label{ms}
Numerical results for the difference $m_t(M_t^2)-M_t$.
The red line represents the $O(\alpha)$ correction, 
the blue line the $O(\alpha) + O(\alpha \alpha_s)$ correction, 
the green  line the $O(\alpha_s)$ correction, 
the magenta line the $O(\alpha_s)+O(\alpha_s^2)$ correction, 
the black line the $O(\alpha_s)+O(\alpha_s^2)+O(\alpha_s^3)$ correction, and
the dark blue line the sum of all these corrections. 
The input parameters are specified in Eq.~(\ref{params}).}
\end{center}
\end{figure}
A numerical estimation of the $O(\alpha_s^4)$ term, given in
Ref.~\cite{kataev-kim}, 
is $\sim -0.02 ~\gv$, which is not included in
Eq.~(\ref{qcd:num}). The analytic result for the EW
corrections at the one-loop order has a more complicated form and may be
found in Refs.~\cite{Yukawa:1,FJ81}.  The two-loop corrections of order
$O(\alpha^2)$ are not yet known. Exploring the results of Ref.~\cite{martin}, 
we estimate it to be of order $ O(1~\gv)$. Another way to
estimate the two-loop contribution follows from results of Ref.~\cite{pole:1} 
and the observation that the largest contribution is coming
from tadpole diagrams:
\begin{eqnarray}
\left\{\frac{m_t(\mu^2)}{M_t} \right\}
\sim 
\left\{\frac{m_t(\mu^2)}{M_t} \right\}_{\rm tadpole}
%= \frac{y_t(\mu^2) v^2(\mu^2)}{\sqrt{2} M_t} 
%= \sqrt{2} \frac{y_t(\mu^2)}{g(\mu^2)} \frac{M_W}{M_t} \sqrt{\frac{m_W^2(\mu^2)}{M_W^2}}
\sim \sqrt{\frac{m_W^2(\mu^2)}{M_W^2}}
= 1 + \frac{1}{2} \delta_{W,\alpha} 
+ \frac{1}{2} \delta_{W,\alpha\alpha_s}
+ \frac{1}{2} \delta_{W,\alpha^2}
-  \frac{1}{8} \delta^2_{W,\alpha}
\;,
\end{eqnarray}
where $\delta_{W,\alpha}$, $\delta_{W,\alpha^2}$, and
$\delta_{W,\alpha_s \alpha}$ are the corrections in the relation
$\frac{m_W^2(\mu^2)}{M_W^2} = 1 + \delta_{W,\alpha} + \delta_{W,\alpha_s \alpha}
+ \delta_{W,\alpha^2}$
and may be found in Ref.~\cite{pole:1}.
This also allows us to estimate the error due to the unknown higher-order corrections, 
which is about $1~\gv$.

\begin{table}[h!]
\centering
\caption{The various contributions to $m_t(M_t^2)-M_t$ in GeV.}
\label{tab:results}
$\begin{array}{|c|c|c|c|c|c|}
%\hline
%\multicolumn{5}{|c|}{m_t(M_t^2)-M_t~[\mbox{GeV}]}         \\   
\hline
M_H~[\mbox{GeV}] & O(\alpha)  & O(\alpha \alpha_s) &
O(\alpha)+O(\alpha \alpha_s) & O(\alpha_s)+O(\alpha_s^2)+O(\alpha_s^3) &
\mbox{total}  \\ \hline
124 & 12.11 & -0.39 & 11.72 & -10.38 & 1.34 \\ \hline
125 & 11.91 & -0.39 & 11.52 & -10.38 & 1.14 \\ \hline
126 & 11.71 & -0.38 & 11.32 & -10.38 & 0.94 \\ \hline
\end{array}$
\end{table}
A detailed comparison of the individual contributions is presented
in Fig.~\ref{ms}.  For a set of experimentally most probable values of
$M_H$ \cite{ATLAS}, $M_H=\{124,125,126\}~\gv$, the
numerical values of the EW and QCD contributions to the difference
$m_t(M_t^2)-M_t$ and their sum are collected in Table~\ref{tab:results}. 
As a result, we observe a large EW correction,
which for the assumed $M_H$ range almost perfectly cancels the QCD
correction. The relationship between $m_t(M_t^2)$ and $M_t$ can be
parametrized in the range displayed in Fig.~\ref{ms} as
\begin{eqnarray}
\left\{ m_t(M_t^2) \!-\! M_t \right\}_{\rm SM}& =& 
\left\{ m_t(M_t^2) \!-\! M_t \right\}_{\rm QCD} + 
\Biggl[0.0664 \!-\! 0.00115\times\left(\frac{M_H}{1~\gv} \!-\! 125\right)
\Biggr]\,M_t \;.
\nonumber\\
&&
\end{eqnarray}
The almost perfect cancellation between the QCD and EW effects for the given
Higgs boson mass is certainly accidental, but must be taken into account in
comparisons with experimental data.
Our calculation shows that the large leading correction, of ${\cal O}(\alpha)$,
to the shift $m_t(M_t^2)-M_t$ is not substantially modified by the
next-to-leading term, of ${\cal O}(\alpha\alpha_s)$.
Radiative corrections beyond the presently known ones are likely to be small
and not to change the observed quenching qualitatively.

\section{Conclusions}

We calculated the shift $m_t(M_t^2)-M_t$ of the top-quark mass in the SM by
strictly taking into account all diagrams generated by the Feynman
rules, including the tadpole ones, as is required to manifestly respect the
Slavnov-Taylor and Ward-Takahashi identities.

SM transition matrix elements of physical processes renormalized
according to the EW on-shell scheme are manifestly devoid of
tadpoles to all orders of perturbation
theory~\cite{Kraus:1997bi}. This has lead to the quite common
practice to set tadpole contibutions to zero. On the other hand, the
tadpoles are gauge dependent, and the mass counterterms are only gauge
independent if the tadpole contributions are included, as may be
observed already at one loop~\cite{Yukawa:1,FJ81}. Also, if tadpoles drop
out from physical quantities, or relations between them, it does not
mean that carrying them along in a calculation would not lead to a
correct result. In contrast, tadpole cancellation may serve as a
useful check of a calculation.

Upon on-shell renormalization, the SM transition matrix elements of
physical processes are gauge-independent functions of the pole masses
and the other renormalized parameters, i.e.\ the couplings and mixing
angles. By finite reparametrizations, these transition matrix elements
may be converted to any other renormalization scheme, in our case to
the \MSb scheme. The relationships between the on-shell paramaters and
the \MSb parameters are gauge independent only if tadpole contributions
are retained. Tadpoles are artifacts of spontaneous symmetry breaking,
where they show up in the Higgs vacuum expectation value, which induces
the masses. Correspondingly, tadpoles affect all mass counterterms,
and only these. The dimensionless gauge and Yukawa couplings as well
as the Higgs self-coupling and their counterterms are not affected.

We advocate to keep the relationships between the on-shell and bare
parameters in the dimensionally regularized theory and their relationships to
the closely related \MSb parameters gauge invariant. Otherwise, the expressions
of the transition matrix elements in terms of the renormalized
parameters acquire artificial gauge dependence, and the choice of
gauge must always be specified, too, whenever the value of a
renormalized parameter extracted from experimental data is
communicated.

This leads us to include the tadpole contribution in the relationship between
the pole mass $M_t$ and the \MSb mass $m_t(\mu^2)$ of the top quark, on which
we focus our attention in this paper.
Assuming the recently-discovered Higgs-like boson to be the missing link of the
SM, then the smallness of its mass $M_H$ renders the positive tadpole
contribution to the difference $m_t(M_t^2)-M_t$ so sizable that it almost
perfectly cancels the familiar $-10$~GeV shift induced by pure QCD
corrections, and it is one of the major purposes of this work to publicize this
intriguing coincidence.
As a welcome consequence of this near-quenching, the theoretical uncertainty
due to scheme dependence in any physical observable that depends on the
top quark mass at leading order is greatly reduced.
In fact, this uncertainty is proportional to the shift $|m_t(M_t^2)-M_t|$ itself,
and is thus reduced by an order of magnitude if $|m_t(M_t^2)-M_t|$ is.
This may easily be understood as follows.
Let $O=f(M)(1+\delta)$ be an $M$-dependent observable with radiative
correction $\delta$ in the on-shell scheme.
In the \MSb scheme, this observable is then given by
$\overline{O}=f(m)(1+\overline{\delta})$ with
$\overline{\delta}=\delta+(M-m)\partial\ln f(m)/\partial m$, and the leading
scheme dependence corresponds to the magnitude of
$O/\overline{O}-1=\delta(M-m)\partial\ln f(m)/\partial m$.

The shift $m_t(M_t^2)-M_t$ is of paramount phenomenological importance for the
combination of different determinations of the top quark mass in ongoing
experiments.
In fact, the value currently extracted by reconstructing the invariant mass of
the top quark decay products is expected to be close to $M_t$ 
\cite{Hoang,pdg,Abazov:2011pta}, while the analysis of the total cross section
of top quark pair production yields a clean determination of $m_t(M_t^2)$
\cite{Moch,pdg,Abazov:2011pta,Langenfeld:2009wd}. 
The EW ${\cal O}(\alpha)$ correction to the $t\overline{t}$
production cross section is available in the on-shell scheme
\cite{Kuhn:2006vh,Beenakker:1993yr}.
In order to consistently incorporate it in the QCD analysis of
Refs.~\cite{Moch,Langenfeld:2009wd}, it needs to be converted to the \MSb
scheme as described above.
This will generate an explicit tadpole contribution in the radiative
corrections to the cross section.
In turn, the scheme dependence will be substantially reduced
because $m_t(M_t^2)$ and $M_t$ almost coincide.

We have analyzed the EW contributions to the running and
scheme dependence of the top quark mass above the $W$ boson threshold,
when $G_F$ can not be treated any longer as a low-energy constant in
one-to-one correspondence with the muon lifetime, but turns into a
running effective parameter. This effect is similar to the running
electromagnetic coupling $\alpha(\mu^2)$, which, however, is strongly
scale dependent right from zero momentum and is sensitive to
non-perturbative hadronic vacuum polarization effects there. Like the
running couplings $g,g',y_f,$ and $\lambda$, also the running of $G_F$
is scheme dependent. In the \MSb scheme, the scale at which $G_F$
effectively starts to run, is not uniquely defined. SM non-decoupling
effects have to be taken into account. In any case, light-fermion
contributions including the one of the bottom quark are tiny.  The
quantitative analysis shows that the main contribution comes from the
matching relation (\ref{matching}), which supplements the RG equation
(\ref{RG}). At low energies, the running of the quark mass is
equivalent to the running of the Yukawa coupling via Eq.~(\ref{eff}) and
by standard QCD corrections.

As the \MSb scheme is a renormalization scheme with mass-independent
anomalous dimensions, mass effects drop out at high energies on account of
their positive canonical mass dimension. This is in contrast to the on-shell
renormalization scheme, where masses are utilized as renormalization
scales, which leads to residual mass effects in the high-energy
asymptotic regime via renormalization effects, with the Callan-Symanzik
equation replacing the \MSb RG equations.

As our focus is on physics at the EW scale, a precise
treatment of mass effects of the heavier SM states $(t, H, Z, W)$ is
mandatory for a precise interpretation of related experimental data.
In particular, for the top quark, which as we know decays before it can
form hadrons, it is not sufficient to take into account QCD
corrections only, as our analysis shows.

In conclusion, for the current value of the Higgs mass, 
$122 < M_H < 128 ~\gv$~\cite{ATLAS}, 
the one-loop EW corrections to $m_t(M_t^2)-M_t$ are
large and have opposite sign relative to the QCD contributions, so that the
total correction is actually small and approximately equal to
$[1 \pm O(1)]~\gv$ (see Table~\ref{tab:results}). 
%The value of $m_t(m_t)$ can be obtained by iteration of
%Eq.~(\ref{matching}) starting from $m_t(M_t^2)$, with the result
%$m_t(m_t) \sim M_t \pm O(3)~\gv.$
As a result, taking into account
EW radiative corrections, besides the QCD ones, reduces the scheme dependence
for EW observables that depend on the top quark mass.

\bigskip

\noindent
\textbf{Acknowledgments:}\\
We are grateful to Leo Avdeev, Dmitri Kazakov, Elisabeth Kraus,
Sergei Mikhailov, Sven Moch, and Oleg Veretin for useful discussions.
This work was supported in part by the German Federal Ministry for
Education and Research BMBF through Grant No.\ 05~H12GUE, by the
German Research Foundation DFG through the Collaborative Research
Center No.~676 {\it Particles, Strings and the Early Universe---The
Structure of Matter and Space-Time} and by the Helmholtz Association
HGF through the Helmholtz Alliance Ha~101 {\it Physics at the
Terascale}. F.J. thanks for support by the EC Program {\it
Transnational Access to Research Infrastructure} (TARI) INFN - LNF,
HadronPhysics3 - Integrating Activity, Contract No.~283286.

%%%%%%%%%%%%%%%%%%%%%%%%%%%%%%%%%%%%%%%%%%%%%%%%%%%%%%%%%%%%%%%%%%%%%%%%%%%%%%%%%%%%%%%%%%%%%%%%%%%%

\end{document}